\documentclass[sigconf,screen]{acmart}

\usepackage{listings}
\usepackage{listings-solidity}
\usepackage{xcolor}
\usepackage{xspace}
\usepackage{pgfplots}
\usepackage{tcolorbox}
\usepackage{multirow}
\usepackage{makecell}
\usepackage{booktabs}
\usepackage{array}

\newcommand{\algname}{\textsc{SmartIntentV2}\xspace}
\newcommand{\algnameold}{\textsc{SmartIntentNN}\xspace}
\newcommand{\algnamebert}{SmartBERT\xspace}

\newcommand{\eg}{\emph{e.g.}\xspace}         
\newcommand{\vs}{\emph{vs.}\xspace}          

\pgfplotsset{compat=1.18}

\begin{document}

\acmYear{2026}\copyrightyear{2026}
\setcopyright{cc}
\setcctype[4.0]{by}
\acmConference[PROMISE '26]{22nd International Conference on Predictive Models and Data Analytics in Software Engineering}{July 5, 2026}{Montreal, QC, Canada}
\acmBooktitle{22nd International Conference on Predictive Models and Data Analytics in Software Engineering (PROMISE '26), July 5, 2026, Montreal, QC, Canada}
\acmDOI{10.1145/3803846.3807464}
\acmISBN{979-8-4007-2584-5/26/07}

\title{Detecting Malicious Intents in Smart Contracts with Pre-trained Programming Language Models}

\author{Youwei Huang}
\affiliation{
	\institution{Independent Researcher}
	\city{Suzhou}
	\country{China}
}
\email{devilyouwei@foxmail.com}

\author{Jianwen Li}
\affiliation{
	\institution{Carnegie Mellon University}
	\city{Moffett Field}
	\state{CA}
	\country{USA}
}
\email{jianwenl@andrew.cmu.edu}

\author{Bin Hu}
\authornote{Corresponding author}
\affiliation{
	\institution{Institute of Computing Technology, Chinese Academy of Sciences}
	\city{Beijing}
	\country{China}
}
\email{hubin09@ict.ac.cn}

\author{Sen Fang}
\affiliation{
	\institution{North Carolina State University}
	\city{Raleigh}
	\state{NC}
	\country{USA}
}

\author{Yao Li}
\affiliation{
	\institution{Macau University of Science and Technology}
	\city{Macao}
	\country{China}
}

\author{Peng Yang}
\affiliation{
	\institution{Institute of Intelligent Computing Technology, Suzhou, CAS}
	\city{Suzhou}
	\country{China}
}

\begin{abstract}
	Malicious developer intents in smart contracts constitute significant security threats to decentralized applications, leading to substantial economic losses.
	Prior work introduced \algnameold, a deep learning model for detecting unsafe developer intents.
	By combining the Universal Sentence Encoder, a K-means clustering-based intent highlighting mechanism, and a Bidirectional Long Short-Term Memory (BiLSTM) network, the model achieved an F1 score of 0.8633 on an evaluation set of 10,000 real-world smart contracts across ten distinct intent categories.

	This paper presents \algname (Smart Contract Intent Neural Network Version 2).
	The primary enhancement is the integration of a BERT-based pre-trained programming language model, which we domain-adaptively pre-train on a dataset of 16,000 real-world smart contracts using a Masked Language Modeling objective.
	\algname retains the BiLSTM-based multi-label classification network for intent detection.
	On the same evaluation set of 10,000 smart contracts, it achieves superior performance with an accuracy of \textbf{0.9789}, precision of \textbf{0.9090}, recall of \textbf{0.9476}, and an F1 score of \textbf{0.9279}, substantially outperforming its predecessor and other baseline models.
	Notably, \algname also delivers a \textbf{65.5\%} relative improvement in F1 score over GPT-4.1 on this specialized task.
	These results establish \algname as a new state-of-the-art model for smart contract intent detection.
\end{abstract}


\maketitle

\section{Introduction}\label{sec:intro}
Smart contracts serve as the foundational infrastructure for decentralized application (DApp) development~\cite{szabo1996smart, antonopoulos2018mastering, ethereum2025smart}, operating on various blockchain platforms, \eg, Ethereum~\cite{buterin2014next, wood2014ethereum} and Binance Smart Chain (BSC)~\cite{cernera2023token}.
These contracts enable decentralized financial services and automate on-chain transactions, fostering a trustless execution environment.
However, the transparency and immutability of smart contracts also introduce significant risks.
Both vulnerabilities and deliberately embedded malicious intents can lead to severe economic losses for DApp users.
While extensive research has been conducted on vulnerability detection in smart contracts~\cite{he2020smart, chu2023survey, chen2025chatgpt}, detecting malicious developer intents remains understudied despite their significant security implications.

To bridge this gap, \algnameold was introduced as a deep learning-based approach for identifying unsafe development intents in smart contracts~\cite{huang2022smartintentnn, huang2025deep}.
It comprises three core components: (1) the Universal Sentence Encoder (USE)~\cite{cer2018universal} for generating contextual embeddings of source code, (2) a K-means clustering-based intent highlighting module to emphasize intent-related features, and (3) a Bidirectional Long Short-Term Memory (BiLSTM) network for multi-label classification across ten distinct categories of unsafe intents.
Evaluations on $10,000$ smart contracts showed that \algnameold achieved an F1 score of 0.8633, with an accuracy of 0.9647, precision of 0.8873, and recall of 0.8406.

In this paper, we introduce \algname (Smart Contract Intent Neural Network Version 2), an enhanced version of this model.
The primary improvement in V2 is the integration of a BERT-based pre-trained language model~\cite{devlin2018bert, liu2019roberta}, specifically CodeBERT~\cite{feng2020codebert}, which replaces USE for embedding generation.
The pre-trained encoder undergoes domain-adaptive pre-training on a corpus of $16,000$ smart contracts using a masked language modeling (MLM) objective to better capture the contextual semantics of smart contract code.
It is subsequently fine-tuned for intent detection via transfer learning, incorporating a BiLSTM-based~\cite{hochreiter1997long, graves2005framewise2} multi-label classification network.
\algname was trained on $16,000$ smart contracts and evaluated on an independent test set of $10,000$ contracts.
The results demonstrate that our model outperforms both traditional baselines and its predecessor, achieving an accuracy of \textbf{0.9789}, precision of \textbf{0.9090}, recall of \textbf{0.9476}, and an F1 score of \textbf{0.9279}.

Furthermore, we compare our model with large language models (LLMs) on the same evaluation dataset, where GPT-4.1 achieves an F1 score of only 0.5606.
In a small-scale test involving 100 smart contracts, GPT-4.1 incurred costs of \textbf{\$2.88} and a latency of \textbf{11,074ms}, demonstrating significantly higher economic and temporal overhead compared to \algname.
These results conclusively establish our model as the state-of-the-art for smart contract intent detection.

Our contributions are as follows:
\begin{itemize}
	\item We present \algname, achieving an F1 score of \textbf{0.9279} for the task of smart contract intent detection.
	\item We introduce \algnamebert, a domain-adapted pre-trained language model specifically designed for smart contract code, publicly available at \textcolor{blue}{\url{https://huggingface.co/web3se/SmartBERT-v2}}.
	\item We have open-sourced the dataset, code, documentation, and models at \textcolor{blue}{\url{https://github.com/web3se-lab/web3-sekit}}.
\end{itemize}

\section{Background}
This section provides the technical background for our work.
We detail the categories of unsafe development intents in smart contracts with representative examples, discuss existing methods for detecting these intents, and review the use of pre-trained language models for code representation.

\subsection{Unsafe Intents in Smart Contracts}\label{intent-categories}
Smart contracts, implemented as Turing-complete programs on blockchain systems, enable the development of DApps.
Solidity is the predominant language for smart contract programming, particularly on platforms like Ethereum and BSC.
While much attention has been given to vulnerabilities inadvertently introduced during development, we argue that intentionally embedded malicious code by developers also constitutes a significant class of contract flaws.
Prior research has identified ten common categories of unsafe development intents in smart contracts: \textbf{Fee}, \textbf{DisableTrading}, \textbf{Blacklist}, \textbf{Reflect}, \textbf{MaxTX}, \textbf{Mint}, \textbf{Honeypot}, \textbf{Reward}, \textbf{Rebase}, and \textbf{MaxSell}~\cite{huang2025deep}.

\begin{figure}[ht]
	\centering
	\begin{lstlisting}[language=Solidity]
function setTxLimit(uint256 amount) external authorized {
    _maxTxAmount = amount;
}

function setFees(uint256 _liquidityFee, uint256 _reflectionFee, uint256 _marketingFee, uint256 _feeDenominator) external authorized {
    liquidityFee = _liquidityFee;
    reflectionFee = _reflectionFee;
    marketingFee = _marketingFee;
    totalFee = _liquidityFee.add(_reflectionFee).add(_marketingFee);
    feeDenominator = _feeDenominator;
    require(totalFee < feeDenominator / 4);
}

function tradingStatus(bool _status) public onlyOwner {
    tradingOpen = _status;
}
    \end{lstlisting}
	\caption{
		A smart contract example exposing the intents of Fee, DisableTrading, and MaxTX.
		The contract address on BSC: 0x20BE792404240f34038d9b20eBCAEbFAA088ee20.
	}
    \Description{Solidity code snippets demonstrating functions that set transaction limits (MaxTX), adjust multiple fee parameters (Fee), and toggle trading status (DisableTrading).}
	\label{fig:intent_example}
\end{figure}

Figure~\ref{fig:intent_example} presents several code snippets from a smart contract that exemplify unsafe intents.
The \texttt{setTxLimit} function allows for unrestricted modification of transaction limits, embodying the \textbf{MaxTX} intent.
This capability can be exploited to manipulate transaction volumes unfairly.
The \texttt{setFees} function facilitates the adjustment of various fees, including liquidity, reflection, and marketing fees, corresponding to the \textbf{Fee} intent. This may lead to unjust transaction costs for users.
Most critically, the \texttt{tradingStatus} function empowers the contract owner to enable or disable trading at will, representing the \textbf{DisableTrading} intent.
This functionality poses a significant threat, as it permits the owner to arbitrarily halt trading operations.
Collectively, these functions reflect unsafe developer intents that have the potential to cause economic harm to users.

\subsection{Smart Contract Intent Detection}
Detecting malicious intents within smart contracts is crucial for safeguarding against potential threats.
\algnameold~\cite{huang2022smartintentnn} was the pioneering model developed to address this challenge.
Built using the TensorFlow.js framework~\cite{abadi2016tensorflow,smilkov2019tensorflow}, it employs a combination of advanced components to enhance detection accuracy: the Universal Sentence Encoder for capturing contextual code representations, a K-means clustering model for highlighting intent indicators, and a BiLSTM network for classifying intents.
The model was trained on a dataset of 10,000 smart contracts and evaluated on a separate, unseen test set of 10,000 contracts.
While it achieved a commendable overall F1 score of 0.8633, significantly surpassing traditional deep learning baselines, it exhibited clear performance bottlenecks.
Specifically, its detection capability was limited for minority-class intents due to data imbalance in the training set.
For instance, the F1 score for the \textbf{MaxSell} intent was only 0.5714, highlighting the need for a more robust model.

\subsection{Pre-trained Language Models}
Pre-trained language models (PLMs) have revolutionized natural language processing with models like BERT~\cite{devlin2018bert} and RoBERTa~\cite{liu2019roberta}.
These models leverage large-scale self-supervised learning to generate robust contextual representations that are adaptable to various downstream tasks via transfer learning, markedly reducing the demand for labeled data.
In the realm of code analysis, models like CodeBERT~\cite{feng2020codebert} extend this paradigm by training on both code and natural language prose.
Our work builds upon this by pre-training CodeBERT specifically on a large dataset of smart contracts, enhancing its ability to discern the subtle semantic nuances pertinent to intent detection.
The resulting model, \algnamebert, is not limited to intent detection. As a domain-specific encoder trained on real-world Solidity code, it produces high-quality function-level embeddings that can benefit a broad spectrum of smart contract analysis tasks, including vulnerability detection, code clone detection, and contract similarity search.
The domain-adaptive pre-training of the encoder on large-scale smart contract corpora is instrumental in equipping \algname with the capacity to model domain-specific syntactic and semantic patterns.

\section{Model}
\begin{figure*}[ht]
	\centering\includegraphics[width=\linewidth]{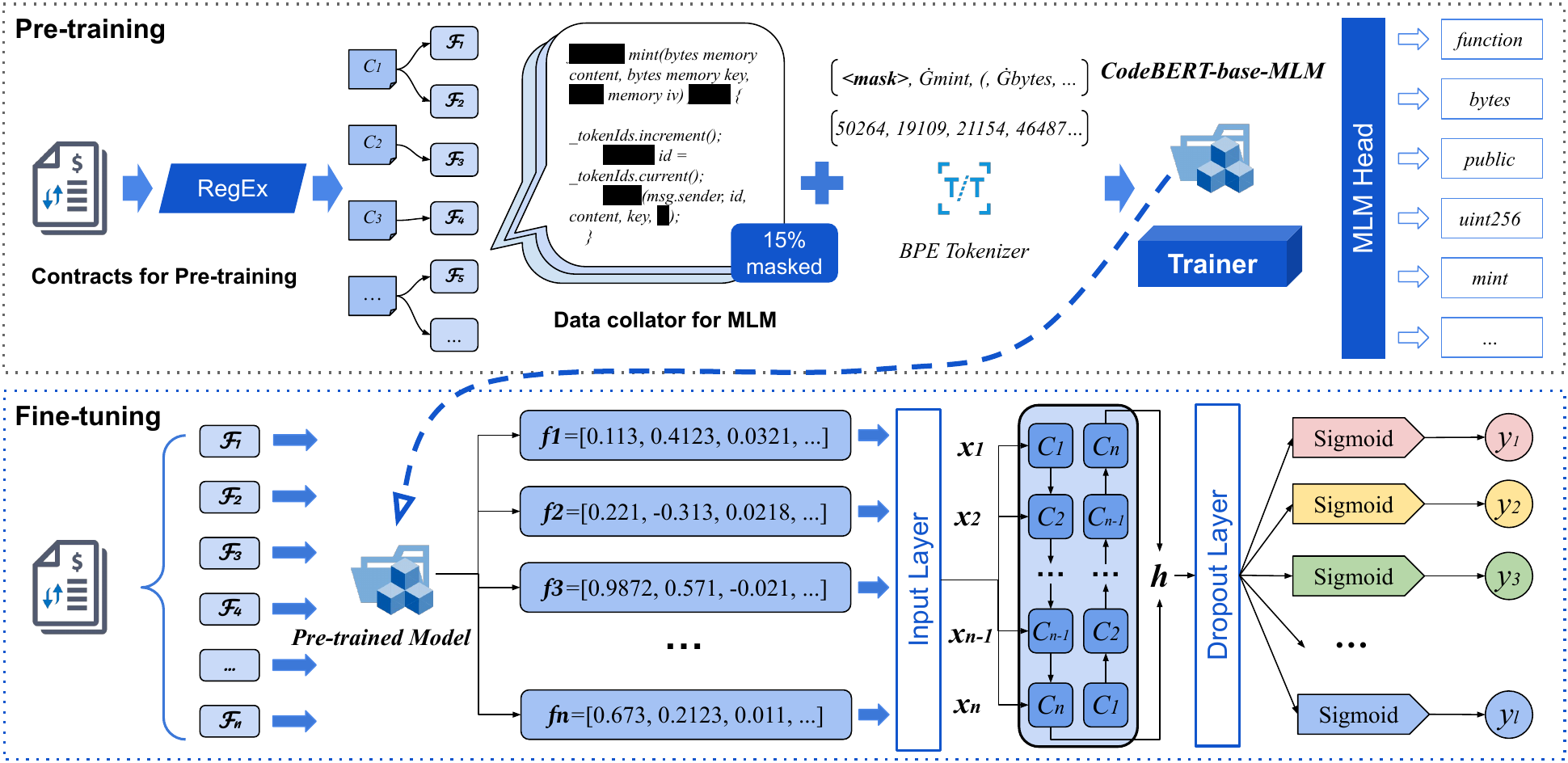}
	\caption{
		Architecture of \algname:
		(i) The initial phase (upper section) involves pre-training on a substantial corpus of smart contracts using the MLM approach, focusing on function-level code;
		(ii) The MLM head is removed to derive an Encoder specifically designed for smart contract functions;
		(iii) A BiLSTM-based network is employed to perform multi-label classification on the smart contracts.
	}
	\Description{The figure shows the architecture of \algname, including pre-training on smart contracts with MLM, removal of the MLM head to obtain an encoder, and a BiLSTM-based network for multi-label classification.}
	\label{framework}
\end{figure*}

\algname comprises \algnamebert, which is a pre-trained model specifically designed for smart contracts, paired with a BiLSTM-based network for multi-label classification.
Figure~\ref{framework} illustrates the complete model architecture of \algname.
The following subsections detail the training process of this pre-trained model and its utilization through transfer learning in downstream tasks for smart contract intent classification.

\subsection{Data Preparation}
\label{sec:data-preparation}
Our dataset originates from an industrial smart contract auditing pipeline, in which contracts deployed on Ethereum and BSC were collected through both automated scanning and manual auditing processes.
Each contract was analyzed at the source code level and labeled for the presence or absence of the ten unsafe intent categories described in Section~\ref{intent-categories}. A single contract may carry zero, one, or multiple labels.

\textbf{Dataset snapshot and scope.}
The auditing pipeline operates continuously.
As of December 2025, the corpus has grown to over 40,000 annotated contracts.
All experiments in this paper were conducted on a \emph{fixed snapshot} taken when the full corpus stood at approximately 40,000 entries.
After rigorous quality control, including cross-verified annotations and removal of ambiguous or duplicated contracts, we retained 30,000 contracts and partitioned them as follows:
\begin{itemize}
	\item \textbf{Training set (16,000 contracts):} used for both \algnamebert domain-adaptive pre-training and \algname classification training. Using the same set for both stages ensures there is no overlap with the evaluation set and that the encoder's pre-training domain is fully aligned with the downstream task.
	\item \textbf{Evaluation set (10,000 contracts):} an independent held-out set used exclusively for all intent-detection experiments, consistent with the configuration in \algnameold.
	\item \textbf{Validation set (4,000 contracts):} disjoint from both the training and evaluation sets, used during \algnamebert pre-training to monitor MLM loss every 10,000 steps.
\end{itemize}
The remaining approximately 10,000 contracts in the corpus at the time of the snapshot were reserved for future studies and not included in the current experiments.
Because the auditing pipeline is continuously growing, incorporating the full corpus into a single experimental cycle is impractical: newly collected contracts require time for annotation, cross-verification, and quality assurance before they can be reliably used for training or evaluation.
We therefore froze the experimental dataset at 30,000 quality-controlled contracts to ensure reproducibility.
The complete dataset, including all contracts collected up to December 2025, has been open-sourced as part of the project repository listed in Section~\ref{sec:intro}.

Due to limitations imposed by maximum sequence length, it is impractical to input entire smart contract contexts into our model during both training and evaluation phases.
Consistent with the approach in \algnameold, we extract code at the function level, with subsequent data processing also conducted at this granularity.
From the perspective of processing an entire smart contract, this approach effectively treats function-level code as sequential data for analysis.

\subsection{Pre-training for Smart Contracts}
We trained \textbf{\algnamebert}, a pre-trained programming language model specifically tailored for smart contract analysis.
The model is initialized from CodeBERT, leveraging its proven capability in code understanding, and adapted to capture the unique syntactic and semantic patterns of smart contract programming.
The pre-training process is visually depicted in the upper part of Figure~\ref{framework}.

\textbf{Training Data and Preprocessing:}
We curated 16,000 real-world smart contracts, covering a broad spectrum of functionalities and structures in decentralized applications.
The dataset was tokenized at the function level, rather than at the contract level, to prevent exceeding the maximum sequence length of 512 tokens and to enable finer-grained semantic representation of code.

\textbf{Pre-training Objective:}
We adopt the MLM objective to adapt the pre-trained \texttt{CodeBERT-base-mlm}\footnote{\url{https://huggingface.co/microsoft/codebert-base-mlm}} model to smart contract code.
In each function-level sequence, 15\% of tokens are randomly selected and replaced with the \texttt{[MASK]} token.
Given an input sequence $\mathbf{x} = (x_1, x_2, \ldots, x_T)$ of length $T$, let $M \subseteq \{1, 2, \ldots, T\}$ denote the masked positions where $|M| \approx 0.15T$.
The model predicts the original tokens $\{x_i \mid i \in M\}$ from the masked context $\mathbf{x}_{\text{masked}}$.
The MLM loss is defined as:
\begin{equation}
	\mathcal{L}_{\text{MLM}} = -\frac{1}{|M|} \sum_{i \in M} \log P_{\theta}(x_i \mid \mathbf{x}_{\text{masked}})
	\label{eq:mlm_loss}
\end{equation}
where $P_{\theta}(x_i \mid \mathbf{x}_{\text{masked}})$ is the predicted token probability at position $i$, parameterized by model weights $\theta$.

\textbf{Model Architecture and Training Setup:}
Our backbone model is \texttt{CodeBERT-base-mlm}, which is initialized from \texttt{RoBERTa-base} and trained with an MLM objective on source code.
We retain its 12-layer encoder architecture with hidden dimension $d=768$ and 12 self-attention heads.
Domain-adaptive pre-training is performed on smart contract code using the MLM objective.
Training runs for 20 epochs with a per-device batch size of 64 on two NVIDIA A100 80GB GPUs (effective batch size 128), taking approximately 10 hours.
We utilize the \texttt{AdamW} optimizer with a learning rate of $5 \times 10^{-5}$.
Evaluation on an independent held-out set of 4,000 smart contracts, disjoint from both the training and evaluation splits described in Section~\ref{sec:data-preparation}, is performed every 10,000 steps during pre-training.

\textbf{Function-level Representation:}
For each function with input sequence $\mathbf{x} = (x_1, x_2, \ldots, x_T)$, \algnamebert produces contextualized token embeddings
$\mathbf{H} = (\mathbf{h}_1, \mathbf{h}_2, \ldots, \mathbf{h}_T) \in \mathbb{R}^{T \times d}$,
where $T$ is the sequence length and $d=768$ is the hidden dimension. We obtain a function embedding $\mathbf{f} \in \mathbb{R}^d$ via mean pooling:
\begin{equation}
	\mathbf{f} = \frac{1}{T} \sum_{t=1}^T \mathbf{h}_t
	\label{eq:function_embedding}
\end{equation}
Empirically, mean pooling outperforms using the \texttt{[CLS]} token embedding for function representation.

\textbf{Contract-level Representation:}
Given a smart contract containing $N$ functions, we obtain function embeddings $\{\mathbf{f}_i\}_{i=1}^N$ using Eq.~\ref{eq:function_embedding} for each function.
These embeddings are stacked to form the contract-level representation matrix:
\begin{equation}
	\mathbf{X} =
	\begin{bmatrix}
		\mathbf{f}_1 \\
		\mathbf{f}_2 \\
		\vdots       \\
		\mathbf{f}_N
	\end{bmatrix} \in \mathbb{R}^{N \times d}
	\label{eq:contract_matrix}
\end{equation}
where each row $\mathbf{f}_i \in \mathbb{R}^d$ represents the $i$-th function embedding.
This matrix $\mathbf{X}$ serves as input to the downstream multi-label classifier for intent detection.

Through this hierarchical representation learning, \algnamebert produces the contract-level matrix $\mathbf{X}$ for downstream multi-label classification.

\subsection{Multi-label Intent Classification}
We formulate smart contract intent detection as a multi-label binary classification task with $C$ target intents ($C=10$ in our dataset).
Given a contract with $N$ functions, the function embeddings $\{\mathbf{f}_i\}_{i=1}^N$ obtained from \algnamebert are organized into a fixed-length matrix through padding or truncation:
\begin{equation}
	\mathbf{X} = \text{pad}\left(\{\mathbf{f}_i\}_{i=1}^N\right) \in \mathbb{R}^{L \times d}
	\label{eq:padded_matrix}
\end{equation}
where $L=256$ is the maximum sequence length. Contracts with fewer than $L$ functions are padded with zero vectors, while longer contracts are truncated.

The BiLSTM encoder processes the padded matrix to capture inter-function dependencies:
\begin{equation}
	\mathbf{h} = \text{BiLSTM}\left(\mathbf{X}\right) \in \mathbb{R}^{2U}
	\label{eq:bilstm_output}
\end{equation}
where $U=128$ is the hidden size per direction, and $\mathbf{h}$ concatenates the final forward and backward hidden states. A masking mechanism ensures padded positions do not contribute to the computation.

The final classification layer applies dropout regularization and produces per-class probabilities:
\begin{equation}
	\mathbf{z} = \mathbf{W} \cdot \text{Dropout}(\mathbf{h}) + \mathbf{b}, \quad
	\hat{\mathbf{y}} = \sigma(\mathbf{z}) \in (0,1)^C
	\label{eq:classification_output}
\end{equation}
where $\mathbf{W} \in \mathbb{R}^{C \times 2U}$, $\mathbf{b} \in \mathbb{R}^C$, dropout rate is 0.5, and $\sigma(\cdot)$ is the element-wise sigmoid function.

\textbf{Padding and Masking:}
Each contract's function sequence is padded to length $L$ by appending zero vectors in $\mathbb{R}^d$.
Denote this operation as $\mathrm{pad}\left(\{\mathbf{f}_i\}_{i=1}^N\right)$.
The Masking layer with $\texttt{mask\_value} = 0.0$ ensures padded timesteps do not contribute to recurrent computations or loss.

\textbf{Loss Function:}
To address class imbalance and emphasize hard examples, we use the binary focal loss~\cite{lin2017focal}. For a single binary label $y \in \{0,1\}$ with prediction $p = \hat{y}$, the focal loss is:
\begin{equation}
	\mathrm{FL}(p, y) = - \alpha \, y (1 - p)^{\gamma} \log(p) - (1-\alpha)(1 - y) p^{\gamma} \log(1 - p)
	\label{eq:focal_loss}
\end{equation}
where $\gamma \geq 0$ is the focusing parameter and $\alpha \in [0,1]$ balances the importance of positive \vs negative samples.
We adopt the default values $\gamma = 2$ and $\alpha = 0.25$, which are widely used in practice~\cite{lin2017focal}.
The overall training loss over $M$ contracts is the mean per-sample, per-class loss:
\begin{equation}
	\mathcal{L} = \frac{1}{M} \sum_{i=1}^M \sum_{c=1}^C \mathrm{FL}(p_{i,c}, y_{i,c})
	\label{eq:total_loss}
\end{equation}
where $p_{i,c}$ and $y_{i,c}$ denote the predicted probability and ground-truth label for class $c$ on sample $i$.
We implement this loss via TensorFlow's \texttt{BinaryFocalCrossentropy}~\cite{abadi2016tensorflow,smilkov2019tensorflow}.

\textbf{Optimization and Training Regimen:}
The classifier is optimized using Adam with the following hyperparameters during the primary training phase (complete data training):
\begin{itemize}
	\item learning rate $1 \times 10^{-3}$
	\item batch size $S = 200$
	\item number of chunks $B = 80$, processing $B \times S = 16,000$ training samples per epoch
	\item epochs per chunk $E = 100$
	\item dropout rate $p = 0.5$
\end{itemize}
A secondary class-balanced training phase uses a reduced learning rate of $1 \times 10^{-4}$, smaller batch sizes, and per-class balanced sampling strategy to further alleviate class imbalance.
Specifically, we randomly sample 10 instances from each intent class to ensure balanced representation across all $C=10$ intent categories during training.

\subsection{Overall Model Summary}
In summary, \algname integrates pre-training, hierarchical representation learning, and sequence modeling for robust intent detection in smart contracts.
The pipeline operates in three stages:
(\rm{i}) function-level embeddings are derived from \algnamebert, a domain-adapted language model pre-trained with the MLM objective on 16,000 real-world contracts;
(\rm{ii}) these embeddings are aggregated into a contract-level matrix with fixed-length padding and masking, ensuring consistent batch processing while preserving semantic granularity;
(\rm{iii}) a BiLSTM-based encoder captures inter-function dependencies, followed by a sigmoid-based output layer that produces multi-label probabilities across all $C=10$ intent categories.

Key design choices include the use of mean pooling for function embeddings, masking-aware sequence encoding, and binary focal loss for handling class imbalance.
Together, these components enhance robustness against sequence length variability and uneven label distribution.
Building on this architecture, we will evaluate \algname on a held-out set of 10,000 smart contracts to demonstrate its effectiveness in intent detection.

\section{Evaluation}
We first formulate three research questions that guide our evaluation, then describe the metrics and baselines used, and finally present the experimental results.

\subsection{Research Questions}\label{sec:rq}
We investigate the following research questions to evaluate the effectiveness of domain-specific pre-training and training strategies in smart contract intent detection.

\begin{tcolorbox}[
		colback=gray!3!white,
		colframe=gray!60!black,
		boxrule=0.5pt,
		arc=2pt,
		title=\textbf{RQ1: Effectiveness of Domain-Specific Pre-training}
	]
	\textbf{Does domain-specific pre-training on smart contracts improve intent detection performance compared to general-purpose pre-trained models?}
\end{tcolorbox}

\begin{tcolorbox}[
		colback=gray!3!white,
		colframe=gray!60!black,
		boxrule=0.5pt,
		arc=2pt,
		title=\textbf{RQ2: Impact of Class-Balanced Training}
	]
	\textbf{How does class-balanced training affect performance under severe intent class imbalance?}
\end{tcolorbox}

\begin{tcolorbox}[
		colback=gray!3!white,
		colframe=gray!60!black,
		boxrule=0.5pt,
		arc=2pt,
		title=\textbf{RQ3: Contributions of Architectural Enhancements}
	]
	\textbf{Which architectural changes in \algname contribute most to the observed performance improvements over its predecessor?}
\end{tcolorbox}

\subsection{Evaluation Metrics}
We evaluate \algname using four standard metrics: accuracy, precision, recall, and F1 score.
For each intent category $c \in \{1, \ldots, C\}$, let $\mathrm{TP}_c$, $\mathrm{TN}_c$, $\mathrm{FP}_c$, $\mathrm{FN}_c$ denote the true positives, true negatives, false positives, and false negatives over all $N$ evaluation samples. The per-class metrics are defined as:
\begin{align}
	\mathrm{Precision}_c & = \frac{\mathrm{TP}_c}{\mathrm{TP}_c + \mathrm{FP}_c}, \quad
	\mathrm{Recall}_c     = \frac{\mathrm{TP}_c}{\mathrm{TP}_c + \mathrm{FN}_c}                                                    \\
	\mathrm{F1}_c        & = \frac{2 \cdot \mathrm{Precision}_c \cdot \mathrm{Recall}_c}{\mathrm{Precision}_c + \mathrm{Recall}_c}
\end{align}
Accuracy is the proportion of correct predictions for class $c$.
F1 provides a balanced measure of precision and recall, and is particularly informative under class imbalance.

We report both \textit{macro-averaged} and \textit{micro-averaged} results.
Macro-averaging computes each metric independently per class and then takes the unweighted mean across all $C$ classes, treating every class equally regardless of its frequency.
Micro-averaging pools the per-class $\mathrm{TP}$, $\mathrm{FP}$, and $\mathrm{FN}$ counts globally before computing the metric, thereby giving more weight to frequent classes.
For instance:
\begin{align}
	\mathrm{Macro\text{-}F1} & = \frac{1}{C} \sum_{c=1}^{C} \mathrm{F1}_c                                                                                \\
	\mathrm{Micro\text{-}F1} & = \frac{2 \cdot \sum_{c} \mathrm{TP}_c}{2 \cdot \sum_{c} \mathrm{TP}_c + \sum_{c} \mathrm{FP}_c + \sum_{c} \mathrm{FN}_c}
\end{align}
The same aggregation schemes apply to accuracy, precision, and recall.

\subsection{Result Analysis}
We evaluate \algname on detecting ten representative intent categories introduced in Section~\ref{intent-categories}.
The evaluation is conducted on a held-out dataset of $N=10{,}000$ real-world smart contracts, disjoint from the training corpus of $16{,}000$ contracts.

\begin{table}[ht]
	\centering
	\caption{Performance of \algname on ten intent detection categories.}
	\resizebox{\columnwidth}{!}{%
		\begin{tabular}{lcccc}
			\toprule
			\textbf{Category}       & \textbf{Accuracy} & \textbf{Precision} & \textbf{Recall} & \textbf{F1}     \\
			\midrule
			\textbf{Fee}            & 0.9452            & 0.9117             & 0.9639          & 0.9371          \\
			\textbf{DisableTrading} & 0.9753            & 0.8954             & 0.8425          & 0.8681          \\
			\textbf{Blacklist}      & 0.9813            & 0.8580             & 0.9190          & 0.8874          \\
			\textbf{Reflect}        & 0.9930            & 0.9806             & 0.9967          & 0.9886          \\
			\textbf{MaxTX}          & 0.9750            & 0.9610             & 0.9579          & 0.9595          \\
			\textbf{Mint}           & 0.9393            & 0.7232             & 0.8630          & 0.7869          \\
			\textbf{Honeypot}       & 0.9910            & 0.5833             & 0.6875          & 0.6311          \\
			\textbf{Reward}         & 0.9947            & 0.9445             & 0.9752          & 0.9596          \\
			\textbf{Rebase}         & 0.9958            & 0.7000             & 0.8953          & 0.7857          \\
			\textbf{MaxSell}        & 0.9987            & 0.7143             & 0.9677          & 0.8219          \\
			\midrule
			\textbf{Macro-average}  & 0.9789            & 0.8272             & 0.9069          & 0.8626          \\
			\textbf{Micro-average}  & 0.9789            & 0.9090             & 0.9476          & \textbf{0.9279} \\
			\bottomrule
		\end{tabular}%
	}
	\label{table:categories}
\end{table}

\begin{figure}[ht]
	\centering
	\includegraphics[width=\columnwidth]{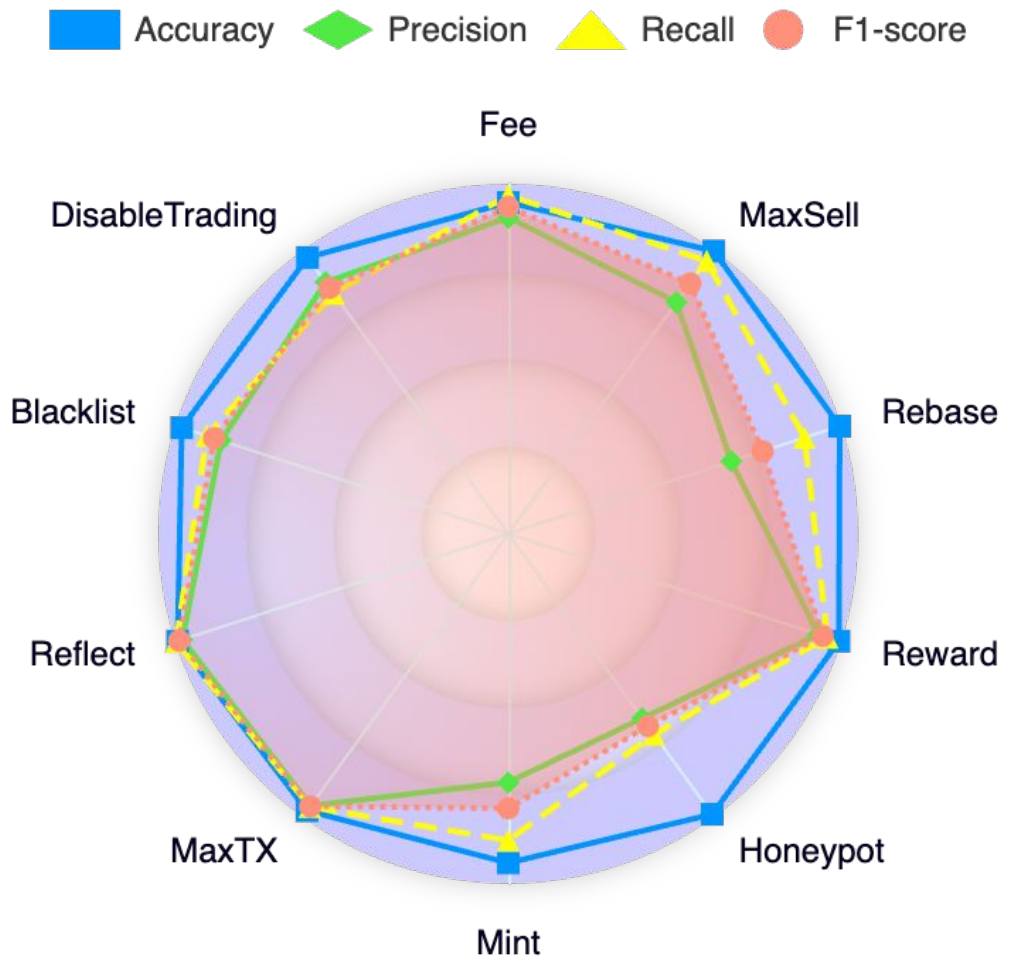}
	\Description{Radar chart showing the performance of \algname on different intent detection tasks, with axes representing various intent categories and values indicating metric scores.}
	\caption{Evaluation of \algname performance on different intent detection tasks.}
	\label{fig:radar}
\end{figure}

As illustrated in Figure~\ref{fig:radar} and detailed in Table~\ref{table:categories}, our model consistently achieves high accuracy, precision, recall, and F1 across all categories.
Notably, the \textbf{Fee}, \textbf{Reflect}, \textbf{Reward}, and \textbf{MaxTX} categories achieve F1 exceeding 0.9, while more challenging and imbalanced intents such as the \textbf{Honeypot}, \textbf{Mint}, and \textbf{Rebase} categories still maintain competitive performance (F1 between 0.63 and 0.79).
This demonstrates that \algname not only excels on majority-class intents but also preserves robustness on minority-class and semantically subtle categories.
The improved balance across intent categories, compared to the predecessor V1, can be attributed to two key enhancements: the incorporation of a secondary class-balanced training strategy that mitigates class imbalance by resampling the training dataset to ensure more equitable representation across all intent categories, and the adoption of binary focal loss during model optimization, which effectively addresses class imbalance by emphasizing hard examples and down-weighting easy negatives, thereby improving performance on minority classes without sacrificing majority class accuracy.

\begin{table}[ht]
	\centering
	\caption{\algname \vs baseline models.}
	\resizebox{\columnwidth}{!}{
		\begin{tabular}{@{}lcccc@{}}
			\toprule
			\textbf{Model}          & \textbf{Accuracy}      & \textbf{Precision}     & \textbf{Recall}        & \textbf{F1}            \\
			\midrule
			\multicolumn{5}{c}{\textbf{\algnameold}}                                                                                     \\
			\midrule
			\textbf{V2 (this work)} & \textbf{\large 0.9789} & \textbf{\large 0.9090} & \textbf{\large 0.9476} & \textbf{\large 0.9279} \\
			V1                      & 0.9647                 & 0.8873                 & 0.8406                 & 0.8633                 \\
			\midrule
			\multicolumn{5}{c}{\textbf{Ablation with Original Pre-trained Models}}                                                      \\
			\midrule
			RoBERTa                 & 0.9693                 & 0.8670                 & 0.9274                 & 0.8962                 \\
			CodeBERT                & 0.9672                 & 0.8516                 & 0.9332                 & 0.8906                 \\
			\midrule
			\multicolumn{5}{c}{\textbf{Other Baselines}}                                                                                \\
			\midrule
			LSTM                    & 0.9172                 & 0.7725                 & 0.5973                 & 0.6737                 \\
			BiLSTM                  & 0.9320                 & 0.7871                 & 0.7200                 & 0.7521                 \\
			CNN                     & 0.9093                 & 0.6922                 & 0.6596                 & 0.6755                 \\
			GPT-3.5-turbo           & 0.8375                 & 0.4135                 & 0.5447                 & 0.4701                 \\
			GPT-4o-mini             & 0.7821                 & 0.3703                 & 0.9240                 & 0.5288                 \\
			GPT-4.1                 & 0.8651                 & 0.4927                 & 0.6501                 & 0.5606                 \\
			\bottomrule
		\end{tabular}%
	}
	\label{table:baselines}
\end{table}

We further compare \algname against its first version and a set of baselines, including traditional deep learning models (\eg, LSTM, BiLSTM, CNN) and LLMs (\eg, GPT-3.5-turbo, GPT-4o-mini, GPT-4.1).
As shown in Table~\ref{table:baselines}, \algname achieves the highest overall accuracy (\textbf{0.9789}), precision (\textbf{0.9090}), recall (\textbf{0.9476}), and F1 (\textbf{0.9279}).
Compared to V1, V2 achieves a \textbf{7.48\%} relative improvement in F1 score, validating the effectiveness of the architectural and training enhancements introduced in this work.
Against conventional neural baselines, \algname delivers substantial relative gains.
For instance, it achieves a \textbf{23.37\%} relative improvement over BiLSTM in F1 score.
The performance gap is even more pronounced when compared to LLM-based baselines, where \algname demonstrates \textbf{75.5\%} and \textbf{65.5\%} relative improvements over GPT-4o-mini and GPT-4.1 respectively in F1, highlighting that task-specific architectures remain highly competitive for domain-specialized classification problems.

This performance gap is largely attributable to the fact that LLMs are pre-trained on general-domain corpora and have not been exposed to smart contract intent detection tasks.
All LLM baselines in this study are evaluated under a zero-shot setting: the prompt contains the natural-language definitions of the ten intent categories and the output format specification, but no labeled input--output examples are provided.
This setting reflects the out-of-the-box capability of general-purpose models on a highly specialized task and establishes a lower bound on LLM performance for smart contract intent detection.
Exploring few-shot prompting with labeled demonstrations and parameter-efficient fine-tuning of LLMs is an interesting direction that we leave for future work (Section~\ref{sec:future-work}).

Beyond performance metrics, we conduct an additional analysis to quantify the economic and temporal costs of smart contract intent detection. We evaluate both \algname and GPT-4.1 on a dedicated test set of 100 smart contracts under concurrent processing conditions. \algname completes the entire task in 2,628ms (2.63s) on a standard PC with negligible computational cost.
In contrast, GPT-4.1 requires 11,074ms (11.07s) with a total consumption of 960,622 tokens (959,801 input tokens and 821 output tokens), averaging 9,606 tokens per request, resulting in approximately \$2.88 in API costs.
This substantial disparity in both time and economic efficiency further reinforces the practical advantages of our specialized approach over general-purpose LLMs for smart contract analysis.

Furthermore, we conduct an ablation study in which \algnamebert is replaced by general-purpose pre-trained models, RoBERTa and CodeBERT, as the encoder for generating smart contract representations, while keeping all other architectural components and training configurations unchanged.
As reported in Table~\ref{table:baselines}, both ablation variants exhibit a noticeable decline in performance: the model with RoBERTa achieves a micro F1 of 0.8962, and with CodeBERT, 0.8906, both substantially lower than the 0.9279 attained by \algname with \algnamebert.
This result underscores the critical role of domain-adaptive pre-training, as \algnamebert's exposure to large-scale smart contract corpora enables it to capture domain-specific syntactic and semantic patterns that are not well represented in general-purpose models.
\algname with \algnamebert consistently outperforms all ablation baselines across all metrics.

\subsection{Answers to Research Questions}

\begin{table*}[ht]
	\centering
	\caption{Ablation study of different backbone models and training strategies for smart contract intent detection.
	Results are reported under both macro- and micro-averaged evaluation metrics.}
	\label{table:ablation_rq}
	\resizebox{\textwidth}{!}{%
		\begin{tabular}{llcccccc}
			\toprule
			\textbf{Backbone Model} & \textbf{Training Strategy} & \textbf{Avg.} & \textbf{Acc.} & \textbf{Prec.} & \textbf{Rec.} & \textbf{F1} \\
			\midrule
			\multirow{4}{*}{\algnamebert}
			 & Original Distribution      & Macro & 0.9751 & 0.8718 & 0.6744 & 0.7125 \\
			 & Original Distribution      & Micro & 0.9751 & 0.9270 & 0.8962 & 0.9113 \\
			\cmidrule(lr){2-7}
			 & Class-Balanced Training    & Macro & \textbf{0.9789} & 0.8272 & 0.9069 & 0.8626 \\
			 & Class-Balanced Training    & Micro & \textbf{0.9789} & \textbf{0.9090} & \textbf{0.9476} & \textbf{0.9279} \\
			\midrule
			\multirow{4}{*}{CodeBERT}
			 & Original Distribution      & Macro & 0.9697 & 0.7594 & 0.6545 & 0.6852 \\
			 & Original Distribution      & Micro & 0.9697 & 0.9017 & 0.8847 & 0.8931 \\
			\cmidrule(lr){2-7}
			 & Class-Balanced Training    & Macro & 0.9672 & 0.7135 & 0.9022 & 0.7735 \\
			 & Class-Balanced Training    & Micro & 0.9672 & 0.8516 & 0.9332 & 0.8906 \\
			\midrule
			\multirow{4}{*}{RoBERTa}
			 & Original Distribution      & Macro & 0.9679 & 0.7267 & 0.6547 & 0.6699 \\
			 & Original Distribution      & Micro & 0.9679 & 0.8813 & 0.8960 & 0.8886 \\
			\cmidrule(lr){2-7}
			 & Class-Balanced Training    & Macro & 0.9693 & 0.7492 & 0.8771 & 0.8026 \\
			 & Class-Balanced Training    & Micro & 0.9693 & 0.8670 & 0.9274 & 0.8962 \\
			\bottomrule
		\end{tabular}%
	}
\end{table*}

We now revisit the three research questions posed in Section~\ref{sec:rq} in light of the experimental results.

\textbf{Answer to RQ1.}
Table~\ref{table:ablation_rq} shows that \algnamebert consistently outperforms general-purpose pre-trained models across all evaluation settings.
Under class-balanced training, \algname with \algnamebert achieves a micro-averaged F1 of \textbf{0.9279}, exceeding CodeBERT (0.8906) and RoBERTa (0.8962).
The advantage is more pronounced under macro-averaged metrics, where \algnamebert reaches an F1 of \textbf{0.8626}, compared to 0.7735 and 0.8026 for CodeBERT and RoBERTa, respectively.
These results indicate that domain-specific pre-training enables \algnamebert to better capture smart contract–specific syntax and semantics that are underrepresented in general programming language corpora.

\textbf{Answer to RQ2.}
As shown in Table~\ref{table:ablation_rq}, class-balanced training substantially improves performance across all backbone models, particularly under macro-averaged metrics.
For \algname with \algnamebert, the macro F1 score increases from 0.7125 to \textbf{0.8626}, corresponding to a \textbf{21.07\%} relative improvement.
Similar trends are observed for RoBERTa (0.6699 to 0.8026, +19.8\%) and CodeBERT (0.6852 to 0.7735, +12.89\%).
These gains primarily stem from improved recall on minority intent categories, which are emphasized by macro-averaged evaluation.
This confirms that class-balanced training is critical for mitigating real-world intent imbalance in smart contracts.

\textbf{Answer to RQ3.}
Compared to V1, \algname achieves a \textbf{7.48\%} relative improvement in F1 score (0.9279 \vs 0.8633).
This gain can be attributed to three key enhancements:
(1) \textbf{Domain-Adaptive Encoder}: Replacing USE with \algnamebert provides richer contextualized representations tailored to smart contract code;
(2) \textbf{Class-Balanced Training}: The secondary training phase significantly improves minority intent detection without degrading majority-class performance;
(3) \textbf{Binary Focal Loss}: This loss function further mitigates class imbalance by emphasizing hard-to-classify samples.
Together, these improvements lead to more robust and balanced intent detection, establishing \algname as a new state-of-the-art approach.

\section{Threats to Validity}

This section discusses various threats to the validity of our study, covering both internal and external factors that could affect our results.
These include data labeling inaccuracies, model and hyperparameter optimization, data imbalance, dataset coverage, and adaptability to evolving attack patterns.

\subsection{Internal Validity}

\textbf{Ground-Truth Accuracy:}
Although we employed both automated pattern matching and manual audits by domain auditors to label the dataset, inaccuracies in multi-label annotations may persist.
We conducted iterative cross-verification to reconcile inconsistent labels and performed additional reviews on randomly sampled subsets to validate annotation quality.
We have also open-sourced our dataset to facilitate further industrial and academic review.
Contracts deemed ambiguous or highly disputed were omitted from inclusion in both the training and evaluation sets.
Despite these measures, minor imperfections in labeling may still exist, affecting the internal validity of our findings.

\textbf{Model Selection and Hyperparameters:}
In our study, we selected \texttt{CodeBERT-base-mlm} as the initialized model for training \algnamebert, acknowledging that the model's architecture and parameter settings may impact subsequent outcomes.
To ensure robust performance, we conducted extensive ablation testing, including experiments with BERT and RoBERTa variants to fine-tune different versions of \algnamebert.
For the downstream classification task, we tested various models, such as LSTM, CNN, and basic feedforward neural networks, along with different hyperparameters, including LSTM units, dropout rates, and normalization layers.
From these numerous experiments, the current model configuration was selected.
However, it is important to note that this setup may not represent the absolute optimum, posing an internal threat.
Thus, it remains open for long-term optimization and refinement.

\subsection{External Validity}

\begin{figure}[ht]
	\centering
	\includegraphics[width=\columnwidth]{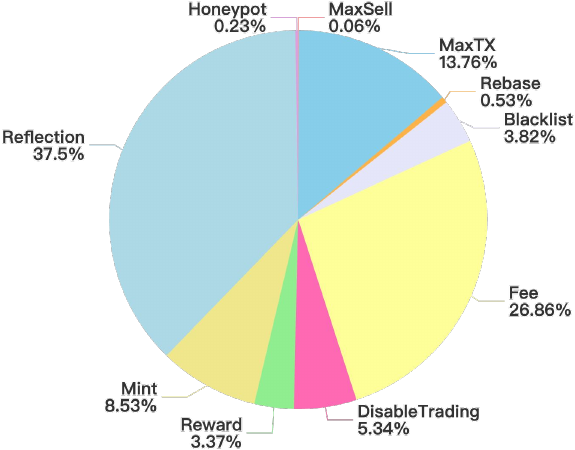}
	\Description{Pie chart showing the distribution of ten intent types, highlighting long-tail minority categories under 1\%.}
	\caption{Distribution of intent types in the dataset.}
	\label{fig:intent_pie}
\end{figure}

\textbf{Data Imbalance:}
As depicted in Figure~\ref{fig:intent_pie}, certain intent categories, such as \textbf{Rebase} and \textbf{Honeypot}, account for less than 1\% of our dataset, reflecting their infrequent occurrence in real-world mainnet code.
Even though a class-balanced second training phase was employed using randomized sampling, this threat persists due to the inherent scarcity of these data samples.
This scarcity suggests that these intents are indeed less prevalent in actual development, thereby naturally reducing their risk profile.
We are committed to continually collecting smart contract data from public blockchain mainnets to incrementally include more of these underrepresented samples.
This ongoing effort will facilitate further training and optimization of our model to better account for these low-frequency intents.

\textbf{Dataset Coverage:}
Our dataset primarily comprises open-source smart contracts from public blockchains like Ethereum and BSC, mainly using Solidity code, although it includes a small number of Vyper contracts~\cite{buterin2018vyper,vyper2025latest}.
There is a lack of data from private and consortium blockchains, which poses a threat to the effectiveness of our method.
We are working on enhancing the dataset with more Vyper contract data.
For non-public and non-open-source contracts, access authorization from the respective platforms is required before they can be incorporated into the training pipeline.

\textbf{Emerging Attacks:}
As the development of smart contracts expands, new categories of malicious intents are likely to emerge, posing potential threats to our model's detection capabilities.
To address this challenge, our neural network model is designed to be both extensible and retrainable.
We can swiftly update our dataset with new samples and conduct extensive training to adapt the model to these evolving threats.
\algname also benefits from its modular architecture; \algnamebert, as a pre-trained model, remains unchanged and does not require retraining.
By focusing on modifying the downstream neural network layers and utilizing a small amount of new data for training, we can effectively identify these unsafe intents, maintaining robustness against emerging attack vectors.

\section{Related Work}
We review prior studies related to our work along three directions: (i) pre-trained models for natural and programming languages, which provide foundations for smart contract analysis; (ii) smart contract vulnerability detection, focusing on external risks; and (iii) malicious behaviors and developer intents, reflecting internal risks.

\subsection{Pre-trained Models for Natural Language and Code}
Pre-trained language models have fundamentally advanced representation learning for both natural language and source code.
In NLP, models such as BERT~\cite{devlin2018bert} and RoBERTa~\cite{liu2019roberta} provide contextualized embeddings that have become standard across downstream tasks.
In the code domain, models including CodeBERT~\cite{feng2020codebert}, CodeT5~\cite{wang2021codet5}, CodeT5+~\cite{wang2023codet5+}, and CuBERT~\cite{kanade2020learning} adapt pre-training objectives to programming languages, enabling tasks such as code search, summarization, clone detection, and defect detection.

More recently, large language models (LLMs) have been explored for smart contract analysis.
Smart-LLaMA-DPO~\cite{yu2025smart} enhances explainability in vulnerability detection through reinforcement learning, while SCALM~\cite{li2025scalm} applies prompting and retrieval strategies to identify insecure coding practices.

Our work is orthogonal to these approaches.
Unlike general-purpose language models and broadly trained code models, we adopt domain-specific pre-training tailored to smart contracts, enabling more precise modeling of contract-specific semantics.
Moreover, in contrast to LLM-based methods that emphasize generative reasoning, our encoder-only architecture is lightweight, controllable, and readily integrable with task-specific neural components such as BiLSTMs.

\subsection{Smart Contract Vulnerability Detection}
Most existing research on smart contract security focuses on vulnerability detection, targeting unintentional defects introduced during development.
Early approaches rely on formal verification, symbolic execution, and static analysis, exemplified by tools such as Oyente~\cite{luu2016making}, Mythril~\cite{mueller2017framework}, ZEUS~\cite{kalra2018zeus}, Securify~\cite{tsankov2018securify}, and SmartCheck~\cite{tikhomirov2018smartcheck}.
{\AE}GIS~\cite{ferreira2020aegis} further extends this line by providing runtime shielding against malicious control and data flows.

With the adoption of machine learning, vulnerability detection has evolved toward learned representations.
SaferSC~\cite{tann2018towards} employs sequential neural models, while ContractWard~\cite{wang2020contractward}, DR-GCN, and TMP~\cite{zhuang2020smart} leverage graph-based learning.
Recent work explores transfer and contrastive learning~\cite{sendner2023smarter, chen2024improving}, as well as LLM-based vulnerability analysis~\cite{chen2025chatgpt, yu2025smart, li2025scalm}.

Despite these advances, the dominant focus remains on bugs, vulnerabilities, and insecure coding practices.
In contrast, our work addresses unsafe developer intents, capturing intentional and high-risk behaviors that cannot be fully characterized by vulnerability-level analysis alone.

\subsection{Malicious Developer Intents and Behaviors}
Internal security risks arise from malicious behaviors deliberately embedded by developers.
Prior studies typically focus on specific threat categories. HoneyBadger~\cite{torres2019art} detects honeypots through symbolic execution, while ``Trade or Trick''~\cite{xia2021trade} and TrapdoorAnalyser~\cite{huynh2025programming} identify scam tokens in decentralized exchanges~\cite{lehar2025decentralized}.
Other work addresses Ponzi schemes~\cite{hu2022scsguard}, backdoors~\cite{ma2023pied}, and rug-pull risks~\cite{zhou2024stop, kalacheva2025detecting, huynh2025serial}, with some employing topological data analysis~\cite{fan2022smart}.

While effective for specific threats, these approaches are largely threat-centric and lack a unified abstraction of developer intent.
Recent work~\cite{huang2025deep} formalized unsafe developer intent categories and demonstrated intent detection using deep learning.
Building on this foundation, the present work advances intent-level representation learning through domain-specific pre-training, enabling a more systematic and proactive form of smart contract security analysis.

\section{Future Work}\label{sec:future-work}
Based on our current study, we identify three directions for advancing smart contract malicious intent detection capabilities.

\textbf{Few-shot and Fine-tuned LLM Comparison.}
Our current LLM baselines are evaluated in a zero-shot setting.
A natural next step is to conduct a comprehensive few-shot evaluation (e.g., 5-shot and 10-shot with demonstrations sampled from the training set) and to explore parameter-efficient fine-tuning of open-weight LLMs on smart contract intent detection data.
Such experiments would establish upper-bound LLM performance on this task and provide a more complete picture of the cost--accuracy trade-off between general-purpose and task-specific models.

\textbf{Bytecode-based Intent Detection for Closed-source Contracts.}
A significant limitation of our current approach is its reliance on source code availability, restricting applicability to open-source smart contracts.
Many deployed contracts on mainnet lack verified source code, creating blind spots in intent detection.
Future work will explore training \algnamebert directly on smart contract bytecode to enable detection of malicious developer intents in closed-source contracts.
This extension would substantially broaden our method's practical deployment scope across the entire blockchain ecosystem.

\textbf{Intent Code Localization Capabilities.}
While our model can successfully identify the presence of 10 malicious intent categories, it lacks precise code localization functionality.
Detecting intent types alone is insufficient for practical security auditing—developers and auditors require exact code locations where malicious intents are implemented.
Future research will investigate extending our model with code localization capabilities, potentially through fine-tuning specialized language models to pinpoint specific functions, statements, or code segments responsible for detected malicious intents.

\section{Conclusion}
We present \algname, an enhanced deep learning model for smart contract intent detection.
The primary enhancement is the incorporation of \algnamebert, a BERT-based programming language model pre-trained on 16{,}000 real smart contracts using MLM at the function level.
This model generates contextual embeddings that are processed by a BiLSTM-based multi-label classifier to detect ten categories of unsafe developer intents.
Comprehensive evaluation on 10,000 smart contracts demonstrates that \algname achieves an accuracy of 0.9789, a precision of 0.9090, a recall of 0.9476, and an F1 score of \textbf{0.9279}, representing a 7.48\% relative improvement over \algnameold (V1) and substantially outperforming all baseline methods including LLMs.
These results establish \algname as the new state-of-the-art model for smart contract intent detection.

\begin{acks}
	This work was supported by the National Key Research Program of China (Grant No. 2021YFF0703800).
\end{acks}

\balance
\bibliographystyle{ACM-Reference-Format}
\bibliography{ref}

\end{document}